# Exact scattering and bound states solutions for novel hyperbolic potentials with inverse square singularity

A. D. Alhaidari

*Saudi Center for Theoretical Physics, P. O. Box 32741, Jeddah 21438, Saudi Arabia*

**Abstract**: We use the Tridiagonal Representation Approach (TRA) to obtain exact scattering and bound states solutions of the Schrödinger equation for short-range inverse-square singular hyperbolic potentials. The solutions are series of square integrable functions written in terms of the Jacobi polynomial with the Wilson polynomial as expansion coefficients. The series is finite for the discrete bound states and infinite, but bounded, for the continuum scattering states.

## 1. Introduction

In quantum mechanics, a physical system and its interaction with the surroundings is described by a wavefunction, which is a solution of the wave equation that contains a potential functions that models the structure and dynamics of the system. Therefore, solutions of the wave equation (e.g., the Schrödinger, the Dirac, etc.) with as many potential models as possible remained one of the prime interest since the early inception of quantum mechanics. Several methods for obtaining exact solutions of the wave equation were introduced. These include, but not limited to, supersymmetry, shape invariance, group theory, operator algebra, factorization, path integral, point canonical transformation, asymptotic iteration, etc. Some of these are equivalent to each other but most address the same class of exactly solvable potentials. Recently, the Tridiagonal Representation Approach (TRA) was developed to handle a larger class of exactly solvable potentials. It is an algebraic approach inspired by the J-matrix method [1] and based on the theory of orthogonal polynomials and their close association with tridiagonal matrices. A recent short review of the TRA is found in [2].

In this work, we use the TRA to obtain exact solutions for the one-dimensional time-independent Schrödinger equation for novel short-range inverse-square singular hyperbolic potentials. In addition to the wavefunction, we write explicitly the discrete bound states energy spectrum and the continuous energy scattering phase shift. In section 2, we formulate the problem within the TRA and in section 3, we obtain the energy spectrum and phase shift.

## 2. TRA formulation of the problem

In the atomic units $\hbar = m = 1$, the time-independent one-dimensional Schrödinger equation for a point particle of mass *m* under the influence of a potential *V(r)* reads as follows

$$\left[ -\frac{1}{2}\frac{d^2}{dr^2} + V(r) - E \right] |\psi(r)\rangle = 0, \qquad (1)$$

where $r \geq 0$, $E$ is the particle energy and $\psi(r)$ its wavefunction. For bound states, the wave function vanishes at the origin and at infinity, whereas for scattering states it is oscillatory at infinity with a bounded amplitude. Now, we make a transformation to a dimensionless coordinate $x(r) = \cosh(\lambda r)$, where $\lambda$ is a real positive scale parameter. Thus, $x \geq 1$ and Eq. (1) is mapped into the following second order differential equation in terms of $x$

$$\lambda^2 \left[ (x^2 - 1) \frac{d^2}{dx^2} + x \frac{d}{dx} - U(x) + \varepsilon \right] \psi(r(x)) = 0, \tag{2}$$

where $\varepsilon = 2E/\lambda^2$ and $U(x) = 2V(r)/\lambda^2$. In accordance with the TRA, we search for a complete set of square integrable basis $\{\phi_n(x)\}$ to expand the wavefunction in a series as $|\psi(r)\rangle = \sum_n f_n |\phi_n(x)\rangle$, where $\{f_n\}$ are the expansion coefficients. Additionally, we require that the basis carries a tridiagonal matrix representation for the wave operator $J = -\frac{1}{2}\frac{d^2}{dr^2} + V(r) - E$. Consequently, the matrix wave equation becomes a three-term recursion relation for $\{f_n\}$. The basis elements $\{\phi_n(x)\}$ are written in terms of classical orthogonal polynomials whose argument is compatible with the range of the configuration space coordinate (i.e., $x \geq 1$). Moreover, the differential equation of these polynomials must have the same structure of the part $p(x)\frac{d^2}{dx^2} + q(x)\frac{d}{dx}$ as that of the differential wave equation (2). The properties of the Jacobi polynomial shown in Appendix A suggest that we can choose the following set of functions as basis elements

$$\phi_n(x) = c_n (x-1)^\alpha (x+1)^\beta P_n^{(\mu,\nu)}(x), \tag{3}$$

where $\mu > -1$, $\mu + \nu < -2N - 1$, $n \in \{0, 1, 2, ..., N\}$ and $N$ is a non-negative integer. Therefore, $\nu$ must be negative whereas the normalization constant is suggested by the orthogonality relation (A5) as $c_n = \sqrt{\frac{\sin\pi(\mu+\nu+1)}{2^{\mu+\nu+1}\sin\pi\nu}}\sqrt{(2n+\mu+\nu+1)\frac{\Gamma(n+1)\Gamma(n+\mu+\nu+1)}{\Gamma(n+\mu+1)\Gamma(n+\nu+1)}}$. The parameters $\alpha$ and $\beta$ will be determined below by the tridiagonal representation requirement. In terms of the variable $x$, the wave operator $J$ is written using Eq. (2) as follows

$$J = -\frac{\lambda^2}{2}\left[ (x^2-1)\frac{d^2}{dx^2} + x\frac{d}{dx} - U(x) + \varepsilon \right]. \tag{4}$$

Using the differential equation of the Jacobi polynomial (A2), we obtain the following action of the wave operator on the basis elements (3)

$$J\phi_n(x) = \frac{\lambda^2}{2} c_n (x-1)^\alpha (x+1)^\beta \left\{ \left[ x(\mu+\nu-2\alpha-2\beta+1) + (\mu-\nu-2\alpha+2\beta) \right] \frac{d}{dx} \right.$$
$$\left. -\frac{\alpha(2\alpha-1)}{x-1} + \frac{\beta(2\beta-1)}{x+1} - (\alpha+\beta)^2 - n(n+\mu+\nu+1) + U(x) - \varepsilon \right\} P_n^{(\mu,\nu)}(x) \tag{5}$$

Using the differential property of the Jacobi polynomials (A4), this action becomes

$$J\phi_n(x) = -\frac{\lambda^2}{2}c_n(x-1)^\alpha(x+1)^\beta \left\{ 2(n+\mu+\nu+1)\left(\frac{2\alpha-\mu-1/2}{x-1} + \frac{2\beta-\nu-1/2}{x+1}\right)\right.$$

$$\left[\frac{(\nu-\mu)n}{(2n+\mu+\nu)(2n+\mu+\nu+2)}P_n^{(\mu,\nu)} - \frac{(n+\mu)(n+\nu)}{(2n+\mu+\nu)(2n+\mu+\nu+1)}P_{n-1}^{(\mu,\nu)} + \frac{n(n+1)}{(2n+\mu+\nu+1)(2n+\mu+\nu+2)}P_{n+1}^{(\mu,\nu)}\right] \quad (6)$$

$$\left. + \left[\frac{\alpha(2\alpha-1)}{x-1} + \frac{\beta(2\beta-1)}{x+1} + (\alpha+\beta)^2 + n(n+\mu+\nu+1) - U(x) + \varepsilon\right]P_n^{(\mu,\nu)}\right\}$$

To produce a tridiagonal representation, the right side of this equation must be a combination of $\phi_n(x)$ and $\phi_{n\pm 1}(x)$ with constant ($x$-independent) factors. The recursion relation of the Jacobi polynomials (A3) shows that (6) will contain terms inside the curly brackets proportional to $P_n^{(\mu,\nu)}$ and $P_{n\pm 1}^{(\mu,\nu)}$ with constant factors only in one of three cases:

(a) $2\alpha = \mu + \frac{1}{2}$ and $(x+1)U(x) = \frac{2\alpha(2\alpha-1)}{x-1} + A + Bx$. (7a)

(b) $2\beta = \nu + \frac{1}{2}$ and $(x-1)U(x) = \frac{2\beta(2\beta-1)}{x+1} + A + Bx$. (7b)

(c) $2\alpha = \mu + \frac{1}{2}$, $2\beta = \nu + \frac{1}{2}$ and $U(x) = \frac{\alpha(2\alpha-1)}{x-1} + \frac{\beta(2\beta-1)}{x+1} + A + Bx$. (7c)

where $A$ and $B$ are arbitrary real dimensionless constants. The terms with $\alpha$ and $\beta$ on the right side of the equations for $U(x)$ are needed to cancel the corresponding terms in Eq. (6) that destroy the tridiagonal structure. Consequently, the potential function corresponding to the case (7a) read as follows:

$$V(r) = \frac{V_0}{\sinh^2(\lambda r)} + \frac{V_+}{\cosh(\lambda r)+1} = \frac{V_0}{\sinh^2(\lambda r)} + \frac{V_+/2}{\cosh^2(\lambda r/2)}, \quad (8a)$$

where $\mu^2 = \frac{1}{4} + 2V_0/\lambda^2$, $A = 2V_+/\lambda^2$ and to make the potential vanish at infinity we took $B = 0$. This is a short-range singular potential with singularity $r^{-2}$ at the origin, which is coming from the first term only and with strength $V_0/\lambda^2$. Reality requires that $2V_0/\lambda^2 \geq -\frac{1}{4}$. This potential can support scattering states, and if $V_0 - 2V_+ > -\lambda^2/8$ then it can also support bound states as we shall see in the following section. The basis parameter $\nu$, which must be negative, is determined by the number of bound states, which is less than or equal to the basis size $N+1$ and it is constrained by the condition that $\mu + \nu < -2N - 1$. In Appendix B, we show that the symmetry of the three-term recursion relation that results from (6) gives $2\beta = \nu + \frac{3}{2}$. On the other hand, the potential function associated with the case (7b) is:

$$V(r) = \frac{V_0}{\sinh^2(\lambda r)} - \frac{V_-}{\cosh(\lambda r)-1} = \frac{V_0}{\sinh^2(\lambda r)} - \frac{V_-/2}{\sinh^2(\lambda r/2)}, \quad (8b)$$

where $\nu^2 = \frac{1}{4} + 2V_0/\lambda^2$, $A = -2V_-/\lambda^2$ and to make the potential vanish at infinity we had to take $B = 0$. This is also a short-range singular potential with a singularity $r^{-2}$ at the origin of strength $(V_0 - 2V_-)/\lambda^2$. Similarly, reality dictates that $2V_0/\lambda^2 \geq -\frac{1}{4}$. This potential can also support scattering states, and if $V_0 - 2V_- > -\lambda^2/8$ then it can support bound states as well. In fact, $V_0 - 2V_+ = -\lambda^2/8$ is the critical singularity strength for this inverse square potential below which quantum anomalies appear [3-4]. The basis parameter $\mu$, which must be greater than $-1$, is determined by the number of bound states and it is constrained by the condition that $\mu + \nu < -2N - 1$. In Appendix B, we show that the symmetry of the three-term recursion relation resulting from (6) gives $2\alpha = \mu + \frac{3}{2}$.

Finally, the potential function for the last case (7c) has a richer structure and reads as follows:

$$V(r) = \frac{V_- + V_+ \cosh(\lambda r)}{\sinh^2(\lambda r)} + V_1 \cosh(\lambda r) + V_0, \tag{8c}$$

where $\mu^2 = \frac{1}{4} + 2(V_+ + V_-)/\lambda^2$, $\nu^2 = \frac{1}{4} + 2(V_+ - V_-)/\lambda^2$, $A = 2V_0/\lambda^2$ and $B = 2V_1/\lambda^2$. Reality in this case requires that $V_+ \geq |V_-| - \frac{1}{8}\lambda^2$. Without the $V_1$ term, this potential becomes the well-known hyperbolic Rosen-Morse potential, which has a well-established exact solution. Moreover, using the identities: $\cosh 2x = \cosh^2 x + \sinh^2 x$, $\sinh 2x = 2(\cosh x)(\sinh x)$ and $\cosh^2 x - \sinh^2 x = 1$ we can show that potential (8c) is identical to

$$V(r) = \frac{W_+}{\sinh^2(\lambda r/2)} + \frac{W_-}{\cosh^2(\lambda r/2)} + V_1 \cosh(\lambda r) + V_0, \tag{8d}$$

where $W_\pm = \frac{1}{4}(V_+ \pm V_-)$. The first two terms are the hyperbolic Pöschl-Teller potential with half the argument. Now, potential (8d) has been treated recently by Assi, Bahlouli and Hamdan using the TRA [5]. Therefore, we will not investigate this potential here but refer the interested reader to the cited work.

Note that the potential function (8b) is obtained from (8a) by the map $x \to -x$ and $V_+ \leftrightarrow V_-$. Moreover, the associated bases (3) are obtained from each other by the additional parameter exchange $\mu \leftrightarrow \nu$ and $\alpha \leftrightarrow \beta$ along with $x \to -x$. We will use this exchange symmetry below to economize on calculation and search only for the solution of the problem with the potential (8a). Applying the said map to this solution will produce the other solution associated with the potential (8b). Hence, the total map becomes as follows

$$x \to -x, \quad V_+ \leftrightarrow V_-, \quad \mu \leftrightarrow \nu, \quad \alpha \leftrightarrow \beta. \tag{9}$$

## 3. TRA solution of the problem

To obtain the exact solution of the problem, we need to identify all ingredients in the wave function $|\psi(r)\rangle = \sum_n f_n |\phi_n(x)\rangle$. Since the basis elements $\{\phi_n(x)\}$ as given by Eq. (3) are now fully determined as shown above, we only need to find an exact realization for the expansion coefficients $\{f_n\}$. To do that, we substitute each of the two potential functions $U(x)$ along with their associated parameters into Eq. (6). Due to the tridiagonal requirement, the matrix wave

equation $J|\psi\rangle = 0$ becomes a three-term recursion relation for the expansion coefficients that will be solved exactly in terms of orthogonal polynomials. In Appendix B, we obtain these symmetric three-term recursion relations associated with the potentials (8a) and (8b). We find that the two resulting recurrence relations, (B5) and (B6), are equivalent to each other under the parameter map (B7), which is equivalent to the map (9) above. Therefore, we consider only one of them, say (B5) associated with the potential (8a). We identify the orthogonal polynomial associated with the recurrence relation (B5) and use the analytic properties of this polynomial to write the phase shift for the scattering states and energy spectrum for the bound states.

We compare (B5) to the recursion relation of the normalized version of the Wilson polynomial $W_n(z^2;a,b,c,d)$ that reads (see Eq. A7 in [6])

$$z^2 W_n = \left[\frac{(n+a+b)(n+a+c)(n+a+d)(n+a+b+c+d-1)}{(2n+a+b+c+d)(2n+a+b+c+d-1)} + \frac{n(n+b+c-1)(n+b+d-1)(n+c+d-1)}{(2n+a+b+c+d-1)(2n+a+b+c+d-2)} - a^2\right] W_n$$
$$-\frac{1}{2n+a+b+c+d-2}\sqrt{\frac{n(n+a+b-1)(n+c+d-1)(n+a+c-1)(n+a+d-1)(n+b+c-1)(n+b+d-1)(n+a+b+c+d-2)}{(2n+a+b+c+d-3)(2n+a+b+c+d-1)}} W_{n-1} \quad (10)$$
$$-\frac{1}{2n+a+b+c+d}\sqrt{\frac{(n+1)(n+a+b)(n+c+d)(n+a+c)(n+a+d)(n+b+c)(n+b+d)(n+a+b+c+d-1)}{(2n+a+b+c+d-1)(2n+a+b+c+d+1)}} W_{n+1}$$

This normalized version of the Wilson polynomial is written as (see Eq. A6 in [6])

$$W_n(z^2;a,b,c,d) = \sqrt{\left(\frac{2n+a+b+c+d-1}{n+a+b+c+d-1}\right)\frac{(a+b)_n(a+c)_n(a+d)_n(a+b+c+d)_n}{(b+c)_n(b+d)_n(c+d)_n n!}} \; {}_4F_3\left(\begin{matrix}-n,n+a+b+c+d-1,a+iz,a-iz\\a+b,a+c,a+d\end{matrix}\bigg|1\right) \quad (11)$$

where ${}_4F_3\left(\begin{matrix}a,b,c,d\\e,f,g\end{matrix}\bigg|z\right) = \sum_{n=0}^{\infty}\frac{(a)_n(b)_n(c)_n(d)_n}{(e)_n(f)_n(g)_n}\frac{z^n}{n!}$ and $(a)_n = a(a+1)(a+2)...(a+n-1) = \frac{\Gamma(n+a)}{\Gamma(a)}$. It is also required that $\text{Re}(a,b,c,d) > 0$ with non-real parameters occurring in conjugate pairs and that $z \geq 0$. However, the comparison of (B5) to (10) shows that either $\text{Re}(a,b) < 0$ or $\text{Re}(c,d) < 0$ and $z$ is pure imaginary. Thus, we define a nonconventional Wilson polynomial, $\tilde{W}_n(z^2;a,b,c,d)$, as a polynomial of degree $n$ in $z^2$ with $n \in \{0,1,2,...,N\}$, $\text{Re}(a+b) > 0$ or $\text{Re}(a+b) < -2N$, $\text{Re}(a+c) < -N + \frac{1}{2}$ and $\text{Re}(a+d) < -N + \frac{1}{2}$. The corresponding three-term recursion relation for $\tilde{W}_n(z^2;a,b,c,d)$ is as follows

$$z^2 \tilde{W}_n = -\left[\frac{(n+a+b)(n+a+c)(n+a+d)(n+a+b+c+d-1)}{(2n+a+b+c+d)(2n+a+b+c+d-1)} + \frac{n(n+b+c-1)(n+b+d-1)(n+c+d-1)}{(2n+a+b+c+d-1)(2n+a+b+c+d-2)} - a^2\right] \tilde{W}_n$$
$$-\frac{1}{2n+a+b+c+d-2}\sqrt{\frac{n(n+a+b-1)(n+c+d-1)(n+a+c-1)(n+a+d-1)(n+b+c-1)(n+b+d-1)(n+a+b+c+d-2)}{(2n+a+b+c+d-3)(2n+a+b+c+d-1)}} \tilde{W}_{n-1} \quad (12)$$
$$-\frac{1}{2n+a+b+c+d}\sqrt{\frac{(n+1)(n+a+b)(n+c+d)(n+a+c)(n+a+d)(n+b+c)(n+b+d)(n+a+b+c+d-1)}{(2n+a+b+c+d-1)(2n+a+b+c+d+1)}} \tilde{W}_{n+1}$$

Comparing (10) to (12) gives $\tilde{W}_n(z^2;a,b,c,d) = (-1)^n W_n(-z^2;a,b,c,d)$. Thus, $\tilde{W}_n(z^2;a,b,c,d)$ will be written in terms of the hypergeometric function ${}_4F_3\left(\begin{matrix}-n,n+a+b+c+d-1,a+z,a-z\\a+b,a+c,a+d\end{matrix}\bigg|1\right)$. Moreover, comparing (B5) to (12) gives $\{f_n\}$ as the nonconventional Wilson polynomials modulo an overall factor independent of $n$. Since $\tilde{W}_0 = 1$, then we can write $f_n = f_0 \tilde{W}_n$.

Additionally, the comparison gives the polynomial parameters and argument in terms of the physical parameters as follows

$$a = \bar{b} = \begin{cases} \frac{\mu+1}{2} + i\sqrt{\varepsilon} \\ \frac{\nu+1}{2} \end{cases} \qquad c = \bar{d} = \begin{cases} \frac{\nu+1}{2} \\ \frac{\mu+1}{2} + i\sqrt{\varepsilon} \end{cases} \qquad z^2 = \frac{V_0 - 2V_+}{2\lambda^2} + \frac{1}{16} \qquad (13)$$

where, in the comparison process, we have used the following identity, which is valid for all $n$ and real parameters $\{\mu, \nu, \chi\}$

$$\frac{(n+\nu+1)(n+\mu+\nu+1)\left[(2n+\mu+\nu+2)^2 + \chi\right]}{(2n+\mu+\nu+1)(2n+\mu+\nu+2)} + \frac{n(n+\mu)\left[(2n+\mu+\nu)^2 + \chi\right]}{(2n+\mu+\nu)(2n+\mu+\nu+1)} = \\ -\frac{4n(n+\mu)}{2n+\mu+\nu} + \frac{1}{2}\left[1 + \frac{\nu^2 - \mu^2}{(2n+\mu+\nu)(2n+\mu+\nu+2)}\right]\left[(2n+\mu+\nu+2)^2 + \chi\right] \qquad (14)$$

Therefore, we obtain $f_n(\varepsilon, V_0, V_+) = f_0(\varepsilon, V_0, V_+) \tilde{W}_n(z^2; a, \bar{a}, c, \bar{c})$. Normalizability of the wave function for bound states and boundedness of scattering states makes the positive definite weight function for $\tilde{W}_n(z^2; a, b, c, d)$ equal to $[f_0(\varepsilon, V_0, V_+)]^2$ [7-8].

The energy spectrum of the discrete bound states for the potential (8a) is obtained from the spectrum formula of $\tilde{W}_n(z^2; a, b, c, d)$, which is obtained in turn from the spectrum formula of the conventional Wilson polynomial by the map $z^2 \to -z^2$. Now, the spectrum formula of the Wilson polynomial is obtained from its asymptotics ($n \to \infty$) and is given by formula (C11) in Appendix C of Ref. [8]. That formula under the map $z^2 \to -z^2$ becomes $z^2 = (k+a)^2$ or $z^2 = (k+c)^2$ depending on whether $a$ or $c$ equals to $\frac{\mu+1}{2} + i\sqrt{\varepsilon}$, giving

$$\varepsilon_k = -\frac{1}{4}\left(2k + 1 + \mu - \sqrt{\mu^2 - 2A}\right)^2 = -\frac{1}{4}\left(2k + 1 + \sqrt{\frac{2V_0}{\lambda^2} + \frac{1}{4}} - \sqrt{\frac{2V_0 - 4V_+}{\lambda^2} + \frac{1}{4}}\right)^2 \qquad (15)$$

where $k = 0, 1, .., k_{max}$ and $k_{max}$ is the largest integer less than or equal to $\frac{1}{2}(\sqrt{\mu^2 - 2A} - \mu - 1)$. On the other hand, we can evaluate the energy spectrum independently using a numerical procedure that starts by writing the recurrence relation (B5) as the matrix eigenvalue equation $T|f\rangle = \varepsilon R|f\rangle$, where $T$ is the tridiagonal matrix obtained by setting $\varepsilon = 0$ in (B5) and $R$ is the tridiagonal matrix multiplying $-\varepsilon$ in (B5). The basis parameters constraint $\mu > -1$ is satisfied by the assignment $\mu = \sqrt{\frac{1}{4} + 2V_0/\lambda^2}$. On the other hand, the basis parameter $\nu$ is still arbitrary but must have an upper bound as $\nu_{max} = -2N - 1 - \mu$. Therefore, we vary the value of $\nu$ between $\nu_{min}$ and $\nu_{max}$ until a plateau of computational stability of the energy spectrum is reached for a conveniently chosen accuracy. We give the results of such computation in Table 1 showing the rate of convergence of the energy spectrum as the size of the basis increases. The Table also demonstrates an excellent agreement with the exact results obtained from the energy spectrum formula (15). We found that the plateau (range of values of $\nu$ with no significant change in the result within the chosen accuracy) is larger for lower bound states, which is typical for such calculations. Specifically, we found that the plateau of computational stability in the basis parameter $\nu$ for the Table is within the range $\nu \in \left[ -2N - \mu - \frac{3}{2}, -2N - \mu - \frac{11}{2} \right]$ and the values given in the Table correspond to the middle of the plateau. Figure 1 is a plot of the bound states wavefunctions corresponding to Table 1, which are computed using the following series

$$\psi_k(r) = \sqrt{2^{\mu+\nu+1}} \sqrt{\sinh(\lambda r)} \left[ \sinh(\lambda r/2) \right]^{\mu} \left[ \cosh(\lambda r/2) \right]^{\nu+1} f_0(\varepsilon, V_0, V_+) \times \\ \sum_{n=0}^{k} c_n \tilde{W}_n \left( z^2; a, \bar{a}, c, \bar{c} \right) P_n^{(\mu,\nu)}(\cosh(\lambda r)) \tag{16}$$

Now, the phase shift for the continuum scattering states is obtained from the asymptotics ($n \to \infty$) of $\tilde{W}_n(z^2; a, b, c, d)$, which is obtained in turn from the asymptotics of the conventional Wilson polynomial $W_n(z^2; a, b, c, d)$ by the map $z^2 \to -z^2$ (i.e., $z \to \pm iz$). Now, the phase shift for the Wilson polynomial is given by formula (C10) in Appendix C of Ref. [8]. Under the map $z^2 \to -z^2$ that formula becomes

$$\delta(E) = -2 \arg \Gamma \left( \frac{\mu+1}{2} - z + i\sqrt{\varepsilon} \right) \tag{17}$$

Mathematically, this is a non-trivial result because the usual asymptotics of orthogonal polynomials is taken as $n \to \infty$ while keeping the argument of the polynomial and its parameters finite. However, in this case two out of the four parameters of the polynomial also tend to infinity as $n \to \infty$. As seen from (13), those are the parameters that are equal to $\frac{\nu+1}{2}$ since it is required that $-\frac{\nu+1}{2} > N + \frac{\mu}{2}$ while $n \leq N \to \infty$.

## 4. Conclusion

In this work, we used the Tridiagonal Representation Approach to obtain the exact solution of the one-dimensional Schrödinger equation for a short-range inverse-square singular hyperbolic potential. The bound (scattering) state wavefunction is written as a finite (infinite) sum of square integrable basis functions in configuration space. The basis functions are written in terms of the Jacobi polynomials and chosen such that the matrix representation of the wave operator is tridiagonal and symmetric. Consequently, the matrix wave equation becomes a symmetric three-term recursion relation for the expansion coefficients of the wavefunction, which is solved in terms of a modified version of the Wilson polynomial. Using the asymptotics of this polynomial we were able to obtain analytic closed form expressions for the discrete bound states energy spectrum and the continuous energy scattering states phase shift. This study is constitute another demonstration of the advantage of using the TRA as a viable alternative method for the solution of quantum mechanical problems.

## Acknowledgements

This work is partially supported by the Saudi Center for Theoretical Physics (SCTP). We are grateful to Prof. H. Bahlouli for fruitful discussions.

## Appendix A: The Jacobi polynomial

For ease of reference, we list here the basic properties of the version of the Jacobi polynomial that we used in this work. It is defined in the usual way, as follows

$$P_n^{(\mu,\nu)}(x) = \frac{\Gamma(n+\mu+1)}{\Gamma(n+1)\Gamma(\mu+1)} {}_2F_1(-n, n+\mu+\nu+1; \mu+1; \tfrac{1-x}{2}) = (-1)^n P_n^{(\nu,\mu)}(-x). \quad (A1)$$

However, here $n = 0, 1, 2, ..., N$, $\mu > -1$ and $\mu + \nu < -2N - 1$ for $x \geq 1$. It satisfies the following differential equation

$$\left\{ (x^2 - 1)\frac{d^2}{dx^2} + \left[ (\mu + \nu + 2)x + \mu - \nu \right]\frac{d}{dx} - n(n+\mu+\nu+1) \right\} P_n^{(\mu,\nu)}(x) = 0, \quad (A2)$$

It also satisfies the following three-term recursion relation

$$xP_n^{(\mu,\nu)} = \frac{\nu^2 - \mu^2}{(2n+\mu+\nu)(2n+\mu+\nu+2)} P_n^{(\mu,\nu)}$$
$$+ \frac{2(n+\mu)(n+\nu)}{(2n+\mu+\nu)(2n+\mu+\nu+1)} P_{n-1}^{(\mu,\nu)} + \frac{2(n+1)(n+\mu+\nu+1)}{(2n+\mu+\nu+1)(2n+\mu+\nu+2)} P_{n+1}^{(\mu,\nu)} \quad (A3)$$

and the following differential relation

$$(x^2 - 1)\frac{d}{dx} P_n^{(\mu,\nu)} = 2(n+\mu+\nu+1) \left[ \frac{(\nu-\mu)n}{(2n+\mu+\nu)(2n+\mu+\nu+2)} P_n^{(\mu,\nu)} \right.$$
$$\left. - \frac{(n+\mu)(n+\nu)}{(2n+\mu+\nu)(2n+\mu+\nu+1)} P_{n-1}^{(\mu,\nu)} + \frac{n(n+1)}{(2n+\mu+\nu+1)(2n+\mu+\nu+2)} P_{n+1}^{(\mu,\nu)} \right] \quad (A4)$$

The associated orthogonality relation reads as follows [9]

$$\int_1^\infty (x-1)^\mu (x+1)^\nu P_n^{(\mu,\nu)}(x) P_m^{(\mu,\nu)}(x) dx = \frac{2^{\mu+\nu+1}}{2n+\mu+\nu+1} \frac{\Gamma(n+\mu+1)\Gamma(n+\nu+1)}{\Gamma(n+1)\Gamma(n+\mu+\nu+1)} \frac{\sin \pi \nu}{\sin \pi(\mu+\nu+1)} \delta_{nm}, \quad (A5)$$

where $n, m \in \{0, 1, 2, ..., N\}$. Equivalently,

$$\int_1^\infty (x-1)^\mu (x+1)^\nu P_n^{(\mu,\nu)}(x) P_m^{(\mu,\nu)}(x) dx = \frac{(-1)^{n+1} 2^{\mu+\nu+1}}{2n+\mu+\nu+1} \frac{\Gamma(n+\mu+1)\Gamma(n+\nu+1)\Gamma(-n-\mu-\nu)}{\Gamma(n+1)\Gamma(-\nu)\Gamma(\nu+1)} \delta_{nm}. \quad (A5)'$$

## Appendix B: Three-term recursion relations for the expansion coefficients of the wavefunction

Substituting the potential function (8a) along with its associated parameters into (6) and defining the parameter $q = 2\beta - \nu - \tfrac{1}{2}$, we obtain

$$J\phi_n(x) = -\frac{\lambda^2}{2} c_n (x-1)^\alpha (x+1)^{\beta-1} \Big\{ 2q(n+\mu+\nu+1)$$

$$\left[ \frac{(\nu-\mu)n}{(2n+\mu+\nu)(2n+\mu+\nu+2)} P_n^{(\mu,\nu)} - \frac{(n+\mu)(n+\nu)}{(2n+\mu+\nu)(2n+\mu+\nu+1)} P_{n-1}^{(\mu,\nu)} + \frac{n(n+1)}{(2n+\mu+\nu+1)(2n+\mu+\nu+2)} P_{n+1}^{(\mu,\nu)} \right] \quad \text{(B1)}$$

$$+ \left\{ \frac{\mu^2}{2} - \frac{(\nu+q)^2}{2} - A + (x+1)\left[ n(n+\mu+\nu+1) + \tfrac{1}{4}(\mu+\nu+1+q)^2 + \varepsilon \right] \right\} P_n^{(\mu,\nu)} \Big\}$$

Using the three-term recursion relation of the Jacobi polynomial (A3) and writing the result in terms of the basis functions $\{\phi_n(x)\}$, we get after some manipulations the following

$$J\phi_n(x) = -\frac{\lambda^2/2}{x+1} \Big\{ (F_n + qn) D_n \phi_{n+1}(x) + \left[ F_n - q(n+\mu+\nu+1) \right] D_{n-1} \phi_{n-1}(x)$$

$$+ \left[ \frac{\mu^2}{2} - \frac{(\nu+q)^2}{2} - A + 2q(\nu-\mu) \frac{n(n+\mu+\nu+1)}{(2n+\mu+\nu)(2n+\mu+\nu+2)} + F_n(C_n+1) \right] \phi_n(x) \Big\} \quad \text{(B2)}$$

where $C_n = \frac{\nu^2 - \mu^2}{(2n+\mu+\nu)(2n+\mu+\nu+2)}$, $D_n = \frac{2}{2n+\mu+\nu+2}\sqrt{\frac{(n+1)(n+\mu+1)(n+\nu+1)(n+\mu+\nu+1)}{(2n+\mu+\nu+1)(2n+\mu+\nu+3)}}$ and $F_n = \varepsilon + n(n+\mu+\nu+1) + \tfrac{1}{4}(\mu+\nu+1+q)^2$. Now, since the wave operator $J$ is Hermitian then its real matrix representation must by symmetric. Therefore, the term that multiplies $\phi_{n+1}(x)$ in Eq. (B2) must be identical to the one that multiplies $\phi_{n-1}(x)$ but with the replacement $n \to n+1$. This is true only if $q = 1$ making $2\beta = \nu + \tfrac{3}{2}$ and mapping Eq. (B2) into the following

$$J\phi_n(x) = \frac{-\lambda^2/2}{x+1} \Big( \left[ \left(n + \tfrac{\mu+\nu}{2} + 1\right)^2 + \varepsilon \right] D_n \phi_{n+1}(x) + \left[ \left(n + \tfrac{\mu+\nu}{2}\right)^2 + \varepsilon \right] D_{n-1} \phi_{n-1}(x)$$

$$+ \left\{ -\frac{2n(n+\mu)}{2n+\mu+\nu} + \left[ \left(n + \tfrac{\mu+\nu}{2} + 1\right)^2 + \varepsilon \right](C_n+1) - \frac{(\nu+1)^2}{2} + \frac{\mu^2}{2} - A \right\} \phi_n(x) \Big) \quad \text{(B3)}$$

where we have used the identity

$$\frac{2(\nu-\mu)n(n+\mu+\nu+1)}{(2n+\mu+\nu)(2n+\mu+\nu+2)} = -\frac{2n(n+\mu)}{2n+\mu+\nu} + n(C_n+1). \quad \text{(B4)}$$

Now, we insert the wavefunction series $|\psi\rangle = \sum_n f_n |\phi_n\rangle$ in the wave equation $J|\psi\rangle = 0$ and use (B3) to obtain the following symmetric three-term recursion relation for the expansion coefficients of the wavefunction

$$(\tfrac{1}{2}\mu^2 - A) f_n = \left\{ \frac{2n(n+\mu)}{2n+\mu+\nu} - \left[ \left(n + \tfrac{\mu+\nu}{2} + 1\right)^2 + \varepsilon \right](C_n+1) + \frac{(\nu+1)^2}{2} \right\} f_n$$

$$- \left[ \left(n + \tfrac{\mu+\nu}{2}\right)^2 + \varepsilon \right] D_{n-1} f_{n-1} - \left[ \left(n + \tfrac{\mu+\nu}{2} + 1\right)^2 + \varepsilon \right] D_n f_{n+1} \quad \text{(B5)}$$

The potential parameter $A = 2V_+/\lambda^2$ is shown explicitly whereas the other parameter $V_0$ is implicit in $\mu$ as $\mu^2 = \frac{1}{4} + 2V_0/\lambda^2$.

Repeating the same calculation for the potential (8b), we find that $2\alpha = \mu + \frac{3}{2}$ and end up with the following symmetric three-term recursion relation for the expansion coefficients of the wave function

$$\left(\tfrac{1}{2}v^2 + A\right)f_n = \left\{\frac{2n(n+v)}{2n+\mu+v} + \left[\left(n + \tfrac{\mu+v}{2} + 1\right)^2 + \varepsilon\right](C_n - 1) + \frac{(\mu+1)^2}{2}\right\}f_n \qquad (B6)$$
$$+ \left[\left(n + \tfrac{\mu+v}{2}\right)^2 + \varepsilon\right]D_{n-1}f_{n-1} + \left[\left(n + \tfrac{\mu+v}{2} + 1\right)^2 + \varepsilon\right]D_n f_{n+1}$$

where here $A = -2V_-/\lambda^2$ and $v^2 = \frac{1}{4} + 2V_0/\lambda^2$. This relation could, in fact, be obtained from the recursion relation (B5) associated with the potential (8a) by the following map

$$\mu \leftrightarrow v, \quad A \to -A, \quad f_n \to (-1)^n f_n. \qquad (B7)$$

The first two parts of this map are equivalent to the potential parameter exchange $V_+ \leftrightarrow V_-$. The last part of the map has its origin in the potential map (9) given at the end of section 2 where $x \to -x$. Using this map and the parameter exchange $\mu \leftrightarrow v$, the property of the Jacobi polynomial (A1) that reads $P_n^{(v,\mu)}(-x) = (-1)^n P_n^{(\mu,v)}(x)$ is the reason behind the last part of the map (B7).

## Table Caption

**Table 1**: The complete finite bound states energy spectrum (in units of $-\frac{1}{2}\lambda^2$) for the potential (8a) with the parameter values $V_0 = 10$ and $V_+ = -80$ (in units of $\frac{1}{2}\lambda^2$). The Table shows the rate of convergence of the calculation with the basis size $N$. It also demonstrates very good agreement with the exact spectrum given by Eq. (15).

**Table 1**

| n | N = 10 | N = 30 | N = 50 | N = 100 | Exact |
|---|---|---|---|---|---|
| 0 | 19.564814269470 | 19.564814269481 | 19.564814269482 | 19.564814269481 | **19.564814269481** |
| 1 | 11.718388029082 | 11.718388037498 | 11.718388037498 | 11.718388037497 | **11.718388037498** |
| 2 | 5.871959151893 | 5.871961805297 | 5.871961805512 | 5.871961805514 | **5.871961805514** |
| 3 | 2.025073374517 | 2.025533739431 | 2.025535451304 | 2.025535570742 | **2.025535573531** |
| 4 | 0.143562312026 | 0.173190908271 | 0.176531383396 | 0.178285719099 | **0.179109341547** |

## Figure Caption

**Fig. 1**: The un-normalized bound states wavefunctions associated with the potential parameters in Table 1. The horizontal axis is the configuration space coordinate, which is measured in units of $\lambda^{-1}$.

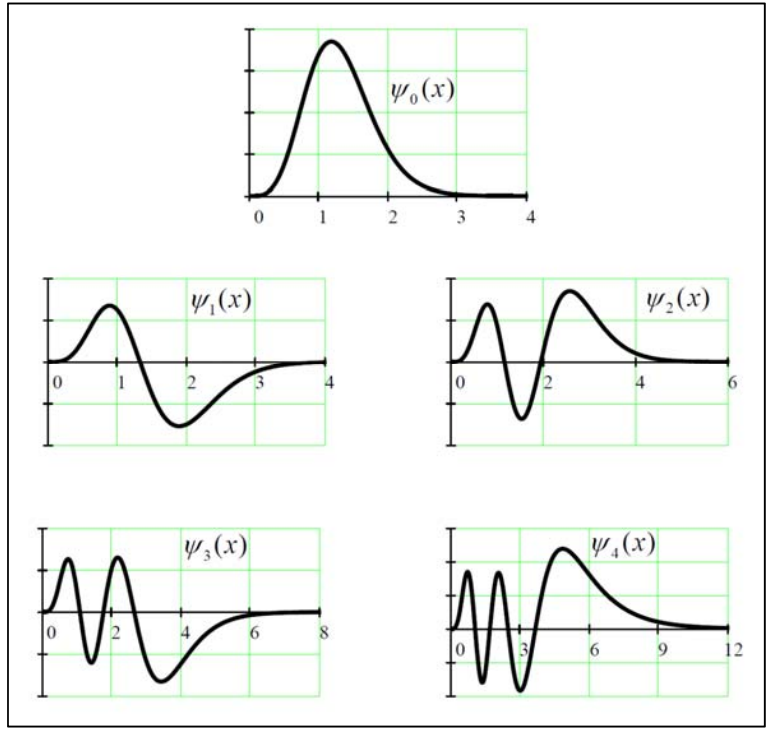

**Fig. 1**